\documentstyle[12pt]{article}
\renewcommand{\theequation}{\arabic{section}.\arabic{equation}}

\newcommand{\lb}[1]{\label{#1}}  
\newcommand{\bb}[1]{\bibitem{#1}}
\def\be{\begin{equation}}
\def\ee{\end{equation}}
\def\ba{\begin{eqnarray}}
\def\ea{\end{eqnarray}} 
\def\ds{\displaystyle}

\def\n{\noindent}
\def\s{\sigma}
\def\e{{\rm e}}
\def\ta{\delta \alpha}
\def\tb{\delta \beta}
\def\tg{\delta \gamma}
\def\tp{\delta \psi}
\def\1p{\delta_1\psi}
\def\0p{\delta_0\psi}
\def\p{^{\prime}}
\def\ff{^{\prime \prime}}
\def\to{\rightarrow}

\begin{document}
\begin{center}

{\Large\bf Electrostatic solutions in Kaluza-Klein theory:\\[5pt]
           geometry and stability}
\medskip

{\bf M.\ Azreg-A\"{\i}nou$^{(a)}$\footnote{e-mail: azreg@taloa.unice.fr},
G.\ Cl\'ement$^{(b)}$\footnote{e-mail: gclement@lapp.in2p3.fr},
C.P. Constantinidis$^{(c)}$\footnote{e-mail: clisthen@cce.ufes.br}, and
J.C. Fabris$^{(c)}$\footnote{e-mail: fabris@cce.ufes.br}}

\end{center}

\medskip

{\small \n $^{(a)}$Girne American University,
Faculty of Engineering, Karmi Campus Karaoglanoglu,
Girne, North Cyprus (via Mersin 10, Turkey)\\ 
$\n ^{(b)}$Laboratoire de Physique Th\'eorique LAPTH (CNRS), B.P.110, 
F-74941 Annecy--le--Vieux cedex, France\\
$\n ^{(c)}$Departamento de F\'{\i}sica, Universidade Federal do
Esp\'{\i}rito
Santo, Vit\'oria, Esp\'{\i}rito Santo, Brazil}

\vskip50mm

\begin{abstract}
We investigate the family of electrostatic spherically symmetric solutions 
of the five-dimensional Kaluza-Klein theory. Both charged and neutral
cases 
are considered. The analysis of the solutions, through their geometrical 
properties, reveals the existence of black holes, wormholes
and naked singularities. A new class of regular solutions is identified.
A monopole perturbation study of all these solutions is carried out,
enabling us to prove analytically the stability of large classes of
solutions. 
In particular, the black hole solutions are stable, while for the regular 
solutions the stability analysis leads to an eigenvalue problem.
\end{abstract}

\newpage
\section{Introduction}

Five--dimensional Kaluza--Klein theory \cite{KK,Jordan}, or sourceless
general
relativity in 4+1 spacetime dimensions (the extra space dimension being
compactified), was historically one of the first unified field theories.
While this simplest higher--dimensional field certainly cannot be
considered as realistic, it nevertheless deserves to be investigated as
the prototype of other multidimensional theories. The static, spherically
symmetric solutions of Kaluza--Klein theory have been obtained
independently by several authors \cite{Leut,DM,CD}. These solutions
include regular black holes, which are generalisations of Schwarzschild
black holes with electric and scalar charges. A systematic investigation
of these black hole solutions was carried out by Gibbons and Wiltshire
\cite{GW}. A class of regular, horizonless charged solutions with wormhole
spatial topology was also identified by Chodos and Detweiler in \cite{CD}.
Apart from these black hole and wormhole solutions, all other solutions
apparently possess naked singularities.

The aim of the present work is to analyze more fully the geometrical
properties of the static spherically symmetric solutions of Kaluza--Klein
theory, as well as to investigate their stability. As far as we know, the
first systematic examination of the stability of these solutions is that
of Tomimatsu \cite{Tom}, which was restricted to electrically neutral
solutions. Tomimatsu concluded that the only stable neutral solution was
the 5--dimensional embedding of the Schwarzschild black hole. The
stability of a class of wormhole solutions was analysed in \cite{stab};
the conclusion was that these were generically stable. This investigation
was generalized in an unpublished work \cite{thesis} to encompass all the
static spherically symmetric solutions. However, our subsequent analysis
of the stability of scalar--tensor black holes \cite{STT} led us to
uncover a flaw in the arguments of \cite{stab,thesis}, which motivated us
to
launch a systematic reinvestigation of the Kaluza--Klein problem.

In the second section of this paper, we recall the construction of the
static, spherically symmetric (in the three ``external'' space dimensions)
solutions to Kaluza--Klein theory. The solutions depend generally on three
parameters $x$, $a$ and $b$, related to the mass, scalar charge and
electrical charge. The analysis of the geometrical invariants and the
geodesics, carried out in section 3, allows the identification of the
different solutions as black holes, wormholes and naked singularities in
the five dimensional space-time.  Some of these naked singularities turn
out to be at an infinite geodesic distance, so that the corresponding
solutions are regular.  The equations for small time--dependent monopole
perturbations of these solutions are then set up in a gauge--independent
fashion in Sect. 4, and decoupled in a special gauge. The transformation
of these perturbations under gauge transformations is also briefly
discussed.

Using this framework, the question of the stability of static solutions to
Kaluza--Klein theory is reduced to an eigenvalue problem. An analytical
investigation of this problem is carried out in Sect. 5. We are able to
prove that two classes of solutions are stable. The first stability class
includes (contrary to Tomimatsu's claim) all the electrically neutral
solutions, while the second stability class includes, among others, all
black hole and extreme black hole solutions. We then discuss two special
cases in which we are able to prove unstability (the second one containing
a stable subcase). In the remaining cases, where we have no analytical
information about the spectrum of eigenmodes, one should resort to
numerical computations to ascertain whether the corresponding solutions
are stable under monopole perturbations.

\section{Dimensional reduction and electrostatic solutions}

The field equations of Kaluza--Klein theory derive from the 5--dimensional
Einstein--Hilbert action
\be
S = -\frac{1}{16\pi G_5}\int d^5x\sqrt{|g_5|}\,R_5\,,
\ee
with the additional assumption that $\partial/\partial x^5$ is a Killing
vector with closed orbits. This last assumption allows the 5--dimensional
metric to be decomposed as
\be\lb{5to4}
ds_5^2 = \bar{g}_{\mu\nu}\,dx^{\mu}\,dx^{\nu} - \e ^{2\psi}(\,dx^5 +
2A_{\mu}\,dx^{\mu})^2\,. 
\ee
The 5--dimensional Einstein equations then reduce to the 4--dimensional
system 
\ba  
& \bar{R}_{\mu}\,^{\nu} & = \; -2\e ^{2\psi}F_{\mu\rho}F^{\nu\rho} +
\e ^{-\psi}\,\bar{D}_{\mu}\,\bar{D}^{\nu}\e ^{\psi}\,, \lb{4Deqs1}\\
& \bar{D}_{\nu}\left(\e ^{3\psi}\,F^{\mu\nu}\right) & = \; 0\,,
\lb{4Deqs2}\\
& \Box\,\e ^{\psi} & = \; - \e ^{3\psi}F_{\mu\nu}F^{\mu\nu}\,, \lb{4Deqs3}
\ea
with $F_{\mu\nu} \equiv A_{\nu,\mu} - A_{\mu,\nu}$.

The first of these equations exhibits a non--minimal coupling of the
scalar field $\e ^{\psi}$ to 4--dimensional gravity. A minimal coupling is
recovered by making the conformal transformation $\bar{g}_{\mu\nu} =
\e ^{-\psi}g^E_{\mu\nu}$ to the Einstein frame, leading to
\be
\label{E}
ds_5^2 = \e ^{-2\varphi/\sqrt{3}}g^E_{\mu\nu}\,dx^{\mu}\,dx^{\nu} -
\e ^{4\varphi/\sqrt{3}}(\,dx^5 + 2A_{\mu}\,dx^{\mu})^2\,,
\ee
with the dilaton field $\varphi = \sqrt{3}\psi/2$. This Einstein frame,
frequently used in the literature \cite{GW}, is not defined when $g_{55} =
- \e ^{2\psi}$ is not negative definite, which will be the case for a
large class of the solutions discussed in this paper. For this reason, we
will use only the string frame defined in (\ref{5to4}). Let us also note
that for $A_{\mu} = 0$ the equations (\ref{4Deqs1})-(\ref{4Deqs3}) reduce
to the Jordan--frame equations of Brans--Dicke theory for $\omega = 0$
\cite{freund}.

The spherically symmetric, electrostatic solutions of Kaluza--Klein theory
have been previously obtained by several authors
\cite{Leut,DM,CD}. To be self--contained, we shall rederive
them here along the lines of \cite{DM}, using the Maison approach to the
dimensional reduction of the higher--dimensional Einstein equations
\cite{Maison}, which we first briefly summarize.  The metric of
$(n+p)$--dimensional spacetime with $p$ commuting Killing vectors may be
parametrized by
\be\lb{n+p}
ds^{\,2} = \lambda_{ab} (dx^{a} + \mbox{\LARGE $a$}^a_i\,dx^i)
(dx^{b} + \mbox{\LARGE $a$}^b_j\,dx^j) + {\tau}^{-1}h_{ij}\,dx^i\,dx^j\,,
\ee
where $i=1,...,n$, $a=n+1,...,n+p$, $\tau = |{\rm det}(\lambda)|$, and the
various fields depend only on the coordinates $x^i$. In our case $n = 3$,
$p = 2$, $x^4$ is the time coordinate and $x^5$ is the Kaluza--Klein
angular coordinate. Using the ($n+p$)--dimensional Einstein equations, the
magnetic--like vector potentials $\mbox{\LARGE $a$}^a_i$ may be dualized
to the scalar twist potentials $\omega_a$ according to
\be
\omega_{a,i} \equiv |h|^{-1/2}\tau\lambda_{ab}h_{ij}\epsilon^{jkl}
\mbox{\LARGE $a$}^b_{j,k}\,.
\ee 
The remaining Einstein equations may then be written as the
$n$--dimensional Einstein--$\s$--model system
\ba
& (\chi^{-1}\chi^{,i})_{;i} & = \; 0 \,,\lb{red1}\\
& R_{ij} & = \; \frac{1}{4}{\rm
Tr}(\chi^{-1}\chi_{,i}\chi^{-1}\chi_{,j})\,,
\lb{red2}
\ea
where the $n$--metric is $h_{ij}$, and $\chi$ is the (anti--)unimodular
symmetric $(p+1)\times(p+1)$ matrix--valued field
\be\lb{chi}
\chi = \left( \begin{array}{ccc}
            \lambda_{ab} + \tau^{-1}\omega_a\omega_b & \tau^{-1}\omega_a\\
            \tau^{-1}\omega_b & \tau^{-1} \end{array} \right)\,.
\ee
Solutions of Eq.\ (\ref{red1}) depending on a single potential $\s(x^i)$
are geodesics 
\be\lb{geotar}
\chi = \eta\,\e ^{A\s}
\ee
of the target space SL($p+1$,R)/SO($p+1$), $\eta$ and $A$ being real
constant matrices (with $|{\rm det}\eta| = 1$, ${\rm Tr}(A) = 0$, $\eta^T
= \eta$, $A^T\eta = \eta A$) and $\s$ a harmonic function,
\be
\nabla^2\s = 0\,.
\ee

Now we specialize to $n = 3$, $p = 2$, and restrict ourselves to
electrostatic solutions, $\omega_a = 0$. If $\s(\infty) = 0$, the metric
(\ref{n+p}) is asymptotically Minkowskian provided
\be
\eta =  \left( \begin{array}{ccc}
                     1 & 0 & 0\\
                     0 & -1 & 0\\
		 0 & 0 & 1
		 \end{array} \right)\,.
\ee
Then the $3 \times 3$ matrix $A$ is block--diagonal and may be
parametrized by
\be\lb{A}
A = \left( \begin{array}{cc}
            N & 0 \\
            0 & -x 
    \end{array} \right)\,, \qquad
N = \left( \begin{array}{cc}
            x-a & b \\
            -b & a 
    \end{array} \right)\,.
\ee
For a spherically symmetric solution with the potential $\s(r)$ normalized
by ${\rm lim}_{(r\to\infty)}\,r\s(r) = -1$, the parameters $x$, $a$ and
$b$ are
related to the physical observables $M$ (ADM mass), $\Sigma$ (scalar
charge) and $Q$ (electric charge) by \cite{spat,Rasheed}
\be
x = 2(M - \Sigma/\sqrt{3})\,, \quad a = -4\Sigma/\sqrt{3}\,, \quad b =
2Q\,.
\ee
We may choose for the spherically symmetric reduced spatial metric the
parametrization 
\be\lb{coord}
-h_{ij}\,dx^i\,dx^j = dr^2 + H(r)\,d\Omega^2\,.
\ee
The harmonic function $\s(r)$ then solves
\be\lb{harm}
\s_{,r} = H^{-1}\,.
\ee
Inserting the ansatz (\ref{coord}) in the reduced Einstein equation
(\ref{red2}), we obtain the system,
\ba
1 - \frac{1}{2}H_{,rr} & = & 0 \,, \cr 
-H^{-1}H_{,rr} + \frac{1}{2}H^{-2}H_{,r}^2 & = & 2\nu^2\,H^{-2}\,,
\ea
with
\be
{\nu}^{2} = \frac{1}{4}\,(x^{2} - y)\,,
\ee
where
\be
\label{Q}
y \equiv {\rm det}N = b^2 + ax - a^2\,.
\ee
This system is solved by
\be
H(r) = r^{2} - {\nu}^{2}\,.
\ee

The form of the function $\s(r)$ obtained by integrating Eq.\
(\ref{harm}) depends on the sign of the constant ${\nu}^{2}$:

a) $y < x^{2}\;({\nu}^{2} > 0)$.

\n In this case,
\be
\s = \frac{1}{2\,\nu}\,\ln\left(\frac{r-\nu}{r+\nu}\right)
\ee
diverges for $r = \nu$. From the form of the function $\tau = \e^{x\s}$,
we see that $r = \nu$ is for $x < 2\nu$ ($x < 0$ or $x \geq 0$, $y < 0$) a
point singularity of the 5--dimensional metric, and for $x \ge 2\nu$ ($x
\ge 0$, $y \ge 0$) a Killing horizon.

b) $y=x^{2}\;({\nu}^{2}=0)$.

\n In this ``extreme'' case the function
\be
\s = - \frac{1}{r}
\ee
diverges for $r = 0$, which is a point singularity if $x < 0$.

c) $y > x^2\;(\nu^2 < 0)$.

\n In this case, which corresponds to the ``Class III'' of \cite{CD}, the
function 
\be\lb{mu}
\s = -\frac{1}{\mu}\,\left(\frac{\pi}{2} -
\arctan\left(\frac{r}{\mu}\right)\right)\,,
\ee
where 
$${\mu}^{2}=-{\nu}^{2} = \frac{1}{4}(y - x^2)\,,$$ 
varies within the finite range $I_{\sigma}=]-\pi/\mu, \,0\,[$ when
$r$ varies from $r \rightarrow -\infty$ to $r \rightarrow +\infty$. We
therefore anticipate that the resulting 5--dimensional metric will be
exempt from singularities.

The full 5--dimensional metric (\ref{n+p}) may be written down once
the matrix $\lambda(r)$, given by (\ref{geotar}) and (\ref{chi}), is
known.
The exponential in (\ref{geotar}) is easily computed with the aid of the
Lagrange interpolation formula \cite{Gan}
\be
f(A) = \sum_i f(z_i) \prod_{j \neq i} \frac{A-z_j}{z_i-z_j}\,.
\ee
Its form depends on the sign of
\be
q^2 \equiv x^2/4 - y\,,
\ee
which determines whether the eigenvalues $z_i$ of the matrix $A$ are real,
degenerate or complex. We obtain in these three cases:
\be\lb{sol}
\begin{array}{lll}
\lambda = & {\ds\frac{1}{q}}\,\e ^{x\s/2}
\left( \begin{array}{cc}
\left({\ds\frac{x}{2}}-a\right)\sinh{q\s} + q\cosh{q\s} & b\sinh{q\s} \\
b\sinh{q\s} & \left({\ds\frac{x}{2}}-a\right)\sinh{q\s} - q\cosh{q\s}
\end{array} \right) 
& \mbox{for}\; y < {\ds\frac{x^2}{4}}\,; \\\\ 
\lambda = & \;\;\e ^{x\s/2}
\left( \begin{array}{cc}
\left({\ds\frac{x}{2}}-a\right)\s + 1 & b\s \\
b\s & \left({\ds\frac{x}{2}}-a\right)\s - 1
\end{array} \right) 
& \mbox{for}\; y = {\ds\frac{x^2}{4}}\,; \\\\
\lambda = & {\ds\frac{1}{p}}\,\e ^{x\s/2}
\left( \begin{array}{cc}
\left({\ds\frac{x}{2}}-a\right)\sin{p\s} + p\cos{p\s} & b\sin{p\s} \\
b\sin{p\s} & \left({\ds\frac{x}{2}}-a\right)\sin{p\s} - p\cos{p\s}
\end{array} \right) 
& \mbox{for}\; y > {\ds\frac{x^2}{4}}\,, 
\end{array}
\\
\ee
\\
\n where $p^2 \equiv y - x^2/4$. In all cases it can be checked that $\tau
= \e ^{x\s}$. We will also use, besides the 5--dimensional metric
(\ref{n+p}), the reduced 4--dimensional fields $\bar{g}_{\mu\nu}$,
$A_{\mu}$ and $\e ^{2\psi}$ obtained from (\ref{5to4}),
which we shall write in the form
\be\lb{canon1}
d\bar{s}^2 = \e ^{2\gamma}\,dt^2 - \e ^{2\alpha}\,d\rho^2 -
\e ^{2\beta}\,d\Omega^2\,, \qquad A_{\mu}dx^{\mu} = V\,dt \,, 
\ee
with
\be\lb{canon2}
\e ^{2\alpha} = -\e ^{-x\s}h_{\rho\rho}\,, \;\; \e ^{2\beta} = H\,\e
^{-x\s}\,,
\;\; \e ^{2\gamma} = \e ^{x\s -2\psi}\,, \;\; \e ^{2\psi} =
-\lambda_{55}\,,
\;\; V = \lambda_{45}/2\lambda_{55}\,.
\ee
If $\rho$ in (\ref{canon1}) is identical with the radial coordinate $r$
defined in (\ref{coord}), $\e ^{2\alpha} = \e ^{-x\s}$.

\setcounter{equation}{0}
\section{Black holes and regular solutions}

\subsection{Case $y < x^2$.}

As observed in the preceding section, the surface $r = \nu$ ($\s \to
-\infty$) where $\tau = |{\rm det}(\lambda)|$ vanishes is a Killing
horizon of the 5--dimensional metric, as well as of the reduced
4--dimensional metric (\ref{canon1}), for $x \ge 0$, $y \ge 0$ \cite{GW}.
However the 5--dimensional metric (\ref{n+p}) and the reduced
4--dimensional fields (\ref{canon2}) are analytical only for $y = 0$.

On the Killing horizon $r = \nu$, the matrix
$\lambda$ reduces for $y = 0$, $x > 0$ to
\be
\lambda_H = \frac{1}{x} 
\left( \begin{array}{ccc}
            a & -b \\
            -b & a-x 
	  \end{array} \right)\,,
\ee
with eigenvalues $0$ and $(2a-x)/x$. If $a \le 0$, then $2a \le 0 < x$, so
that the non--zero eigenvalue $(2a-x)/x$ is negative and the non--null
Killing directions are spacelike, corresponding to an event horizon. The
black--hole solutions of this class \cite{DM}, which includes the
5--dimensional Schwarzschild metric for $a = b = 0$, have been extensively
studied by Gibbons and Wiltshire \cite{GW}.  On the other hand, if $a >
0$, then (from $y = 0$) $x = a - b^2/a < a < 2a$, so that the non--null
Killing directions are timelike, and the Killing horizon is actually a
conical singularity. The neutral solution $b = 0$ ($x = a$)
\be
ds^2 = dt^2 - \frac{r-\nu}{r+\nu}\,(dx^5)^2 - \frac{r+\nu}{r-\nu}\,dr^2 -
(r+\nu)^2\,d\Omega^2
\ee
is the direct product of the time axis by the Euclidean Schwarzschild
metric, and is regular if the period of $x^5$ is $8\pi\nu$ \cite{CD}. The
general $b \neq 0$ solution behaves near $r = \nu$ as the product of the
(2+1)--dimensional metric generated by a spinning point particle
\cite{DJH} by the 2--sphere S$_2$.

In the case $y \neq 0$, it is generally assumed that $r = \nu$ is a
curvature singularity \cite{DM}\cite{GW}. This assumption is based on an
extrapolation from the neutral case $b = 0$. However the extrapolation
goes through only if $y < x^2/4$, in which case the matrix $\lambda$ has
real eigenvalues and can always be diagonalized, so that the
5--dimensional curvature invariants of a charged solution are identical
with those of some neutral solution. On the other hand, if $y \ge x^2/4$,
then the 5--dimensional metric cannot be diagonalized, and we must compute
the metric invariants to ascertain whether they diverge on the Killing
horizon. 

The components of the 5--dimensional curvature tensor may be computed from
the formulas given in \cite{thesis}. Using the parametrization (\ref{A}),
(\ref{coord}) and Eq. (\ref{harm}), we obtain
\ba
& R_{abcd} & = \frac{1}{4}\,\tau H^{-2}\, [(\lambda N)_{ac}(\lambda
N)_{bd}
- (\lambda N)_{ad}(\lambda N)_{bc}]\,, \nonumber \\
& R_{arrb} & = \frac{1}{4}\, H^{-2}\, [\lambda\,(N^2 + (x-4r)N)]_{ab}\,,
\nonumber \\
& R_{a\theta\theta b} & = \frac{1}{4}\, H^{-1}(2r - x)(\lambda N)_{ab}\,,
\\
& R_{r\theta r\theta} & = \frac{1}{4}\, \tau^{-1}H^{-1}(y - x^2 + 2xr)\,,
\nonumber \\
& R_{\theta\varphi\theta\varphi} & = \frac{1}{4}\,
\tau^{-1}\sin^2\theta\,(2x^2 - y - 4xr)\,. \nonumber
\ea
The computation of the 5--dimensional Kretschmann invariant $K_5 =
R^{\alpha\beta\gamma\delta}R_{\alpha\beta\gamma\delta}$ is easily
carried out by tracing powers of the matrix $N$, for instance 
\be
{R^{ab}}_{cd}{R^{cd}}_{ab} = \frac{1}{8}\,\tau^2 H^{-4} \left( [{\rm
Tr}(N^2)]^2 - {\rm Tr}(N^4) \right)\,,
\ee
leading to the compact result
\be\lb{K}
K_5 = \frac{3}{2}\,\tau^2 H^{-4}\,[y^2 + 4\rho xy + 8\rho^2(x^2-y)]\,,
\ee
with $\rho \equiv r - x/2$. 

>From (\ref{K}) we obtain the behaviour of the Kretschmann scalar near the
Killing horizon ($r \simeq \nu$), 
\be
K_5 \simeq \frac{3(x-2\nu)^2(2x^2-3y)}{512\nu^8} \left(
\frac{r-\nu}{2\nu} \right)^{(x-4\nu)/\nu}\,.
\ee 
For $\nu > x/4$ ($y < 3x^2/4$) this diverges on the Killing horizon,
except in the special cases $\nu = x/2$, which corresponds to the $y = 0$
black holes, or $y = 2x^2/3$ \footnote{However the fact that in this case
the Kretschmann scalar vanishes on the horizon does not seem relevant, as
the 5--dimensional metric is not analytical and cannot be extended through
the horizon.}. On the other hand, for $\nu \le x/4$ ($3x^2/4 \le y < x^2$)
the 5--dimensional Kretschmann scalar is finite or vanishes on the Killing
horizon.  The form of the Kretschmann scalar $K_4$ for the reduced
4--dimensional metric (\ref{canon1}) is more involved as it depends on
that of the Kaluza--Klein scalar field $\lambda_{55}$. We have found that
on the Killing horizon $r = \nu$, it likewise diverges for $y < 3x^2/4$
(except for $y = 0$), and is finite or vanishes for $3x^2/4 \le y < x^2$.

Finite curvature invariants often, but not always, signal regularity of
the spacetime geometry. A recent counter--example is that of ``cold black
hole'' solutions \cite{STT,CF}, which have everywhere finite
curvature invariants but, because of a lack of analyticity which prevents
a Kruskal--like extension through the Killing horizon, are generically
singular, except for a discrete set of solutions. In the present case we
will see that the $3x^2/4 \le y < x^2$ solutions are indeed regular,
although
non--analytical on the horizon. They need not be extended through the
Killing horizon because this is at infinite geodesic distance. In other
words, they are already geodesically complete.

The energy integral for geodesic motion in the spherically symmetric
5-dimensional metric (\ref{n+p})-(\ref{coord}) may be written as 
\be\lb{geo5}
\dot{r}^2  + \frac{l^2\tau^2}{r^2-\nu^2} + V(r) = 0 \,, \quad V \equiv
\tau(\eta -\Pi^T\lambda^{-1}\Pi)\,, 
\ee
where $\Pi_4$, $\Pi_5$ and $l$ are constants of the motion proportional to
the energy, electric charge and angular momentum of the test particle,
$\dot{r} \equiv dr/d\rho$ with $\rho$ an affine parameter, and the
(arbitrarily
scaled) integration constant $\eta$ is positive, zero or negative for
timelike, null or spacelike geodesics. Assuming $x^2/4 < y < x^2$, $x >
0$ (the Killing horizon can be shown to be at finite geodesic distance for
$y \le x^2/4$, $x > 0$), we obtain from (\ref{sol}) the effective
potential
\be\lb{pot}
V = \e^{x\s}\,(\eta - \alpha\e^{-x\s/2}\sin(p\s+\beta))
\ee
where the constants $\alpha$ and $\beta$ depend on $\Pi_4$ and $\Pi_5$.

For $l = 0$, $\e^{-x\s/2}$ increases when when the horizon $r = \nu$ ($\s
\to -\infty$) is approached so that, whatever the values of the constants
of the motion, with $\alpha \neq 0$, there is a $\s_1$ such that
\be
\e^{-x\s_1/2}\sin(p\s_1+\beta) = \eta/\alpha\,,
\ee
corresponding to a turning point of the geodesic (reflection on a
potential barrier). The only geodesics which can reach the horizon are
those with $\alpha = 0$ ($\Pi_4 = \Pi_5 = 0$) and $ \eta < 0$, leading to
\be
\rho \sim (r-\nu)^{(4\nu-x)/4\nu}\,.
\ee
It then follows that:

a) If $x^2/4 < y < 3x^2/4$ ($x < 4\nu$), those geodesics
terminate on the horizon (on which, as shown previously, the curvature
invariant $K_5$ diverges). The 5-dimensional spacetime is
geometrically singular, although this singularity is harmless to physical
test particles ($\eta \ge 0$), which will be reflected away before hitting
the singularity (the same can be shown to be true for $y = x^2/4$ if 
$x < 2a$).

b) If $3x^2/4 \le y < x^2$ ($x \ge 4\nu$), the horizon (on which $K_{(5)}$
is finite or vanishes) is at infinite geodesic distance, so that the
5--dimensional spacetime is geodesically complete.
While these 5--dimensional spacetimes are geometrically regular, they
admit closed timelike curves. From (\ref{sol}), $\lambda_{55}$
changes sign periodically, so that all the circles $r = \theta = \varphi =
t =$ const. are timelike in the domains where $\lambda_{55} > 0$. This is
linked with the fact that the reduced 4--dimensional metric $g_{\mu\nu}$
in
(\ref{5to4}) is singular on the spheres $\lambda_{55} = 0$, which separate
4--Minkowskian regions ($\lambda_{55} < 0$) from 4--Euclidean regions
($\lambda_{55} > 0$). We take the view that these 4--dimensional
singularities are artefacts due to the breakdown of 5--to--4 dimensional
reduction, i.e.\ the choice of a bad coordinate system (\ref{5to4}) for
the 5--dimensional geometry, which is perfectly regular in the
parametrization (\ref{n+p}).

\subsection{Case $y = x^2$.}

In this case $\nu = 0$ and $\s = -1/r$. We again assume $x \ge 0$, as the
5--dimensional geometry is obviously singular at $r = 0$ if $x < 0$. In
the
case $x = 0$, then $y = 0$ and $b = \pm a$, leading to the 5--dimensional
``extreme black hole'' metric with flat spatial sections \cite{DM,spat} 
\be
ds^2 = \left(1 + \frac{a}{r}\right)\,dt^2 \mp \,\frac{2a}{r}\,dt\,dx^5 -
\left(1-\frac{a}{r}\right)(dx^5)^2 - dr^2 -r^2\,d\Omega^2\,.
\ee
While from (\ref{K}) the Kretschmann scalar vanishes in this case, the
study of geodesic motion shows that this geometry is actually singular.
The effective potential
\be
V = \eta + {\Pi_5}^2 - {\Pi_4}^2 + (\Pi_4 \pm \Pi_5)^2\,\frac{a}{r}
\ee
is unbounded from below for $a < 0$, in which case all the $l = 0$
geodesics
terminate at the singularity $r = 0$. For $a > 0$, only geodesics with $l
= \Pi_4 \pm \Pi_5 = 0$, $\eta < 0$ terminate at the singularity, which is
thus harmless to physical test particles. 

The general case $x > 0$ is quite similar to the case b) above, the
5--dimensional spacetime being geodesically complete, with an infinite
number of changes of sign of $\lambda_{55}$.

\subsection{Case $y > x^2$.}

Inspection of the 5--dimensional metric (\ref{n+p}), (\ref{sol}) with $\s$
given by (\ref{mu}) shows that for all values of $x$ it is regular
\cite{CD} for $r \in ]-\infty, +\infty[$ , so that the 5--dimensional
geometry is of the Lorentzian wormhole type. However, as pointed out in
\cite{geo} (in the special case $x = 0$ of the symmetrical wormhole),
these wormholes are non-traversable, in the sense that physical
(non--tachyonic) test particles cannot go from one asymptotically flat
region ($r \to +\infty$) to the other ($r \to -\infty$). This would be
possible only if the effective potential $V(r)$ in (\ref{geo5}) was
negative over the whole range of $r$. Here the potential (\ref{pot}) with
$\eta \ge 0$ is necessarily positive over part of this range, as the range
$]-p\pi/\mu, 0[$ of $p\s$ is at least $2\pi$ ($p^2 \ge 4\mu^2$), and a
physical test particle coming from $r \to + \infty$ is always
reflected back to $r \to + \infty$, just as in the case $x^2/4 < y \le
x^2$.

Likewise, 5--dimensional light cones also gradually tumble over when $r$
decreases from $+\infty$ to $-\infty$, leading to the existence of closed
timelike curves. The 5--dimensional metric is asymptotically Minkowskian
at both points at spatial infinity if $p = 2n\mu$ ($y/x^2 = (n^2 -
1/4)/(n^2 - 1)$) with $n$ integer, in which case light cones tumble over
by $n\pi$ between the two points at infinity, and the solution is a
metrical kink of winding number $n$ \cite{kink} (other axisymmetric kink
solutions of Kaluza--Klein theory are discussed in \cite{topo}).

\setcounter{equation}{0}
\section{Small perturbations}

In this section we set up the equations for small time--dependent
spherically symmetric perturbations of the electrostatic solutions. As
shown in \cite{stab}, we may choose for the five--dimensional metric a
parametrization similar to that of (\ref{5to4}),(\ref{canon1}) with metric
functions now depending on time,
\be\lb{timedep}
ds_5^2 = \e^{2\gamma(\rho,t)}\,dt^2 - \e^{2\alpha(\rho,t)}\,d\rho^2 -
\e^{2\beta(\rho,t)}\,d\Omega^2 - \e^{2\psi(\rho,t)}(dx^5 +
2V(\rho,t)\,dt)^2\,.
\ee
The Kaluza--Klein Gauss law (Eq. (\ref{4Deqs2})) may be integrated to
\be
F^{41} = Q\,\e^{-\alpha-2\beta-\gamma-3\psi}\,,
\ee
with $Q$ the conserved electric charge. The remaining time--dependent
spherically symmetric equations (\ref{4Deqs1}) and (\ref{4Deqs3}) then
reduce to the system
\ba
& \e^{-2\beta+\alpha-\gamma-\psi}
(\,\e^{2\beta-\alpha+\gamma+\psi}\,\psi')' 
- \e^{2(\alpha-\gamma)}\,\ddot{\psi} 
= -2Q^2\,\e^{2(\alpha-2\beta-2\psi)}\,, & \lb{td1}\\
& \e^{-2\beta+\alpha-\gamma-\psi}
(\,\e^{2\beta-\alpha+\gamma+\psi}\,\beta')'
- \e^{2(\alpha-\gamma)}\,\ddot{\beta} 
= \e^{2(\alpha-\beta)}\,, & \lb{td2}\\
& \e^{-2\beta+\alpha-\gamma-\psi}
(\,\e^{2\beta-\alpha+\gamma+\psi}\,\gamma')' -
\e^{2(\alpha-\gamma)}\,(\ddot{\alpha} + 2\ddot{\beta} + \ddot{\psi})
= 2Q^2\,\e^{2(\alpha-2\beta-2\psi)}\,, \lb{td3}&\\
& -2\beta'' - \psi'' + \beta'(2\alpha'-3\beta') + 
\psi'(\alpha' - 2 \beta' - \psi') + \e^{2(\alpha-\beta)} 
= Q^2\,\e^{2(\alpha-2\beta-2\psi)}\,, & \lb{td4}\\
& -2\dot{\beta}' - \dot{\psi}' + (2\beta' + \psi')\dot{\alpha} 
+ 2(-\beta' + \gamma')\dot{\beta} + (\gamma' - \psi')\dot{\psi} = 0 \, & 
\lb{td5}
\ea
with $' = \partial/\partial\rho$, $\dot{} = \partial/\partial t$.
These equations are not all independent. The first three dynamical
equations correspond respectively to Eq. (\ref{4Deqs3}) and to the
${\bar{R}_2}^2$ and ${\bar{R}_4}^4$ components of Eq. (\ref{4Deqs1}),
while the last two constraint equations correspond to
$({\bar{R}_4}^4-{\bar{R}_i}^i)/2$ and $\bar{R}_{41}$ respectively.

Now we linearize the metric fields in (\ref{timedep}) around the static 
background fields in Eqs. (\ref{canon1}), 
\be\lb{pert}
\psi (\rho,t) = \psi (\rho) + \tp (\rho,t)\,,\;\mbox{etc}\,,
\ee
where $\tp\,,\ta\,,\tb\,,\tg$ are small perturbations. The linearization
of equations (\ref{td1})--(\ref{td5}) leads to the differential system for
the perturbations, which has been simplified by using the static equations
of motion, and choosing a ``harmonic'' background coordinate system in
which the background fields $\psi\,,\alpha\,,\beta\,,\gamma$ are related
by
\be
\label{bg}
\alpha - 2\beta - \gamma - \psi = 0
\ee
(which amounts to choosing the radial coordinate $\rho = \sigma$, with
$h_{\rho\rho} = H^2$):
\ba
\lefteqn{\tp \ff - \psi \p \ta \p + 2\psi \p \tb \p + \psi \p \tg \p + 
\psi \p \tp \p} \nonumber \\
& & - \e^ {2(2\beta + \psi)}\delta\ddot{\psi} - 2\psi \ff (\ta - 2\tb -
2\tp)
= 0\,, \label{S} \\
\lefteqn{\tb \ff - \beta \p \ta \p + 2\beta \p \tb \p + \beta \p \tg \p +
\beta \p \tp \p} \nonumber \\
& & - \e^ {2(2\beta + \psi)}\delta\ddot{\beta} - 2\beta \ff (\ta - \tb)
= 0\,, \label{R22} \\
\lefteqn{\tg \ff + \gamma \p (- \ta \p + 2\tb \p + \tg \p + \tp \p)
- \e^ {2(2\beta + \psi)}(\delta\ddot{\alpha}} \nonumber \\
& & + 2\delta\ddot{\beta} +
\delta\ddot{\psi}) + 2\psi \ff (\ta - 2\tb - 2\tp) = 0\,, \label{R00} \\ 
\lefteqn{2\tb \ff + \tp \ff - (2\beta \p + \psi \p)\ta \p + 2(\beta \p -
\gamma \p)\tb \p + (\psi \p - \gamma \p)\tp \p} \nonumber \\
& & - (2\beta \ff + \psi \ff) \ta + 2(\beta \ff + \psi \ff) \tb + 2\psi
\ff
\tp = 0\,, \label{H} \\
\lefteqn{2\delta\dot{\beta}\p + \delta\dot{\psi}\p = (2\beta\p + \psi\p)
\delta\dot{\alpha} + 2(-\beta\p + \gamma\p)\delta\dot{\beta} + (\gamma\p - 
\psi\p)\delta\dot{\psi} \,.} 
\label{R01}
\ea
The last two constraint equations are not independent, as the  time
derivative of (\ref{H}) may be seen (using $\gamma'' + \psi'' = 0$, which
follows from  (\ref{canon2})) to be identical with the space derivative
of (\ref{R01}).

The perturbations $\tp$, $\ta$, etc. are defined only up to a change of
coordinates preserving the form of the metric (\ref{timedep}), so that we
still have the ``gauge freedom" to choose coordinates for the perturbed
spacetime by imposing a supplementary relation between these
perturbations. The separation of the linearized equations
(\ref{S})--(\ref{R01}) is simpler in the gauge where the perturbations
$\1p$, $\delta_1\alpha$, etc. are constrained by the gauge
condition\footnote{In the Einstein frame (\ref{E}) this gauge condition
reads simply $\delta_1\beta_E = 0$.}
\be\label{j1}
2\delta_1\beta + \1p =0\,.
\ee
The integration of the constraint equation (\ref{R01}) then leads to the
relation 
\be\label{ta}
\delta_1\alpha = \frac{\psi \p -\beta \p}{\psi \p +2\beta \p}\,\1p\,.
\ee
With the help of the unperturbed field equations and Eq. (\ref{j1}), 
Eqs. (\ref{S}), (\ref{R22}) combined together according to: $\beta \p
\times$(\ref{S}) $-$ $\psi \p \times$(\ref{R22}), lead to the wave
equation for $\1p$
\be\lb{wave}
\1p \ff - \e ^{2(2\beta + \psi)}\delta_1\ddot{\psi} 
- 6\frac{F'}{F^2}\,\1p = 0\,,
\ee
with
\be\lb{F}
F  \equiv 2/\psi' + 1/\beta'\,.
\ee
>From a  solution $\1p(\s,t)$ to this master equation, the other
perturbations in the gauge (\ref{j1}) may  be obtained by using Eqs.
(\ref{j1}), (\ref{ta}), and the equation
\be
\label{tg}
\delta_1\gamma\p = 2\,\frac{\psi \p -\beta \p}{\psi \p +2\beta \p}\,
\1p\p - \delta_1\alpha\p\,,
\ee
obtained by adding Eqs. (\ref{S}) and (\ref{R22}) according to
(\ref{S}) $+$ $2\times$(\ref{R22}), and using, first Eq. (\ref{j1}) to
simplify the obtained equation, then Eq. (\ref{ta}) to cancel
$\psi \ff\,,\beta \ff$.

One can also gauge transform these perturbations  $\1p$,
$\delta_1\alpha$, etc. to obtain the perturbations $\tp$, $\ta$, etc. in a
generic gauge, by carrying out the coordinate transformation $(t_1,\rho_1)
\to (t,\rho)$. For instance the scalar field $\psi$ can be linearized
around its static value in  both coordinate systems,
\ba
\psi(t,\rho) & = & \psi(\rho) + \tp(t,\rho) + \cdots \nonumber \\
& = & \psi(\rho_1) + \1p(t_1,\rho_1) + \cdots \nonumber \\
& = & \psi(\rho_1) + \psi'(\rho_1)\delta_1\rho(t_1,\rho_1) + 
\tp(t,\rho) + \cdots \lb{gt1} \,,
\ea
where we have linearized the coordinate transformation according to $\rho
\simeq \rho_1 + \delta_1\rho(t_1,\rho_1)$. To first order Eq. (\ref{gt1})
leads to
\be\lb{gt2}
\tp = \1p - \psi'\,\delta_1\rho\,.
\ee
Eliminating $\delta_1\rho$ between (\ref{gt2}) and a similar equation for
the scalar perturbation $\tb$, and using the gauge condition (\ref{j1}),
we obtain the relation
\be\lb{gt3}
\beta'\tp - \psi'\tb = \frac {1}{2}\,(2\beta' + \psi')\1p\,.
\ee
>From this last relation it can be shown that our wave equation
(\ref{wave}) is identical with the scalar wave equations (18) or (20) of
\cite{stab} (see also \cite{thesis}) derived in the gauge $\delta_2\beta =
0$, the relation between the Kaluza--Klein scalar perturbations,
\be\lb{gt4}
\1p = \frac{2\beta'}{2\beta' + \psi'}\,\delta_2\psi\,,
\ee
leading to the identification with the notations of \cite{stab},
\be\lb{gt5} 
\1p \equiv -\Phi/2 \equiv -fR/2\,,
\ee
with $f \equiv \e^{-2\psi}\,2\beta'/(2\beta' + \psi')\,$.

\setcounter{equation}{0}
\section{Stability}

We now address the question of stability of electrostatic
solutions against radial perturbations. Given stationary perturbations of
the form $\1p (\s,t)=\1p (\s)\,\e ^{i\Omega t}$, etc., we
search for non-trivial real solutions to the equation obtained from 
(\ref{wave}) by assuming $\Omega$ imaginary, $\Omega =-ik$ ($k>0$),
\be\lb{master}
\1p\ff - (k^2\,\e ^{2(2\beta + \psi)} + 6F'/F^2)\,\1p = 0\,,
\ee
where $F$ is defined in (\ref{F}).  The existence of such solutions,
satisfying some physical boundary conditions, means that any initially
small perturbation will grow exponentially in time, thus the background
solution is unstable. Conversely, the background solution is stable if all
the eigenvalues $\Omega$ are real.

The effective potential function $U(\s)$---the coefficient of
$\1p$ in the second term of Eq. (\ref{master})---has double poles
at the zeroes $\s_i$ of $F$, i.e. the roots of $2\beta \p + \psi \p =0$. 
Near such possible poles, the effective potential behaves as
\be
U(\sigma) \;\simeq\;
-\frac{12{\beta'_i}^2}{(2\beta''_i+\psi''_i)}\,\frac{1}{(\sigma-\s_i)^2}
\;=\; \frac{2}{(\sigma-\s_i)^2} \,,
\ee
with $\beta'_i = \beta'(\s_i)$, etc., from which follows the behaviour
\be
\1p \simeq \frac{C_1}{\sigma-\s_i} + C_2(\sigma-\s_i)^2
\ee
($C_1$ and $C_2$ integration constants). So $\1p$ generically has
poles at $\sigma = \s_i$. However such poles turn out to be
spurious, being induced by the gauge fixing $2\delta_1\beta +
\1p = 0$. Indeed, it follows from Eq. (\ref{gt3}) that perturbations $\tp$
and $\tb$  which are regular in a generic gauge lead, in the gauge
(\ref{j1}), to a $\1p$ with poles at $\s_i$.  Conversely these poles can
be removed by transforming to another gauge.  Similarly, from Eq.
(\ref{gt4}) spurious poles also occur in the gauge $\delta_2\beta = 0$ at
the zeroes  of $\beta'$ \cite{STT,stab,thesis}\footnote{In Refs.
\cite{stab} and \cite{thesis}, solutions for which such poles occurred
were argued to be stable, on the basis that generic perturbations
$\delta_2\psi$ were unbounded; however this argument is not gauge
invariant.}.  

These spurious poles being discarded, divergences of the perturbation
$\1p$ can only occur at the two ends of the range $I_{\s}$ of $\s$, and
must be excluded by the choice  of appropriate boundary conditions. We
shall adopt here, as physically reasonable from the 5--dimensional
general--relativistic point of view, the boundary conditions previously
used in scalar--tensor theories \cite{STT}, which state that the relative
perturbations of the background fields must be finite at the boundary.
Since our functions
$\delta_1\alpha,\,\delta_1\beta,\,\delta_1\gamma,\,\1p$ are relative
perturbations of the background metric fields in (\ref{5to4}), these
boundary conditions read
\be\lb{bc}
|\delta_1\alpha| < \infty\;\;\;,\;\;\;|\delta_1\beta| <
\infty\;\;\;,\;\;\;
|\delta_1\gamma| < \infty\;\;\;,\;\;\;|\1p| <
\infty\;\;\;,\;\;\;\mbox{for}\;\;
\s \;{\in} \; I_{\s}
\ee
(the last of these boundary conditions was termed ``strong''  boundary
condition in Ref. \cite{STT}). The  boundary conditions (\ref{bc}) lead
after gauge transformation to similar boundary conditions in a generic
gauge.

Eq. (\ref{master}) behaves at spatial infinity
($r \to +\infty ,\,\s \to 0_{-}$) as
\be
\label{comp1}
\1p \ff - \frac{k^2}{\s ^4}\,\1p \simeq 0\,,
\ee
for all values of the parameters ($x,\,y,\,a$), leading to the asymptotic
behaviour
\be\lb{comp2}
\1p \simeq \s \,\left(c_1 \e ^{k/\s} + c_2 \e ^{-k/\s}\right)\,.
\ee
Our boundary conditions are satisfied by choosing $c_2 =0$. 

A number of possible behaviours of Eq. (\ref{master}) can occur at the
lower end $\s_{\rm min}$ of the range $I_{\s}$, depending on the values of
the parameters ($x,\,y,\,a$). In most cases \cite{thesis},
 the corresponding general
asymptotic solution will be a linear combination of a bounded and an
unbounded solution with coefficients $c_3$ and $c_4$. Then the
perturbation 
will remain bounded near $\s_{\rm min}$ provided $c_4$ is fixed equal 
to zero. However, owing to the scale invariance of Eq. (\ref{master}), 
the two conditions $c_2 = c_4 = 0$ can be satisfied simultaneously only
for special values of the parameter $k^2$, i.e. we have an eigenvalue
problem. In the absence of the knowledge of the exact solution in the
whole interval, it is impossible in general to solve analytically this
eigenvalue problem, and thus to establish with certainty the stable 
or unstable character of the background solution. 
However there are two parameter ranges, corresponding to neutral
solutions, and to a preferred set of charged solutions, for which
stability
can be proved analytically by straightforward arguments. We will examine 
these cases in the next two subsections, and discuss in a third subsection 
two special cases for which unstability can be proved analytically.  

\subsection{Stability of neutral solutions}

In the parameter space ($x,\,y,\,a$), the domain of neutral solutions is
obtained from (\ref{Q}) by setting $b=0$. Since $a$ is real, this leads to
the portion of the surface

\be\lb{ns}
a^2 - xa + y = 0\,,
\ee
for which $y\leq x{^2}/4$ ($q^2 >0$, with $q = |x/2 - a|$). As recalled in
the Introduction,  neutral  Kaluza--Klein theory corresponds to
Brans--Dicke theory with $\omega = 0$. Therefore the proof, given in
\cite{STT} that all  static spherically symmetric solutions to
Brans--Dicke  theory are stable under radial pertutbations carries over
to the present case. For completeness, we sketch here this  proof in the
Kaluza--Klein setting.

In the neutral case, the static scalar equation (\ref{td1}) reduces to
$\psi'' = 0$ in the coordinate system (\ref{bg}), so that the linearized
equation (\ref{S}) immediately decouples from the other equations in the
harmonic gauge
\be\label{j0}
\delta_0\alpha - 2\delta_0\beta - \delta_0\gamma - \delta_0\psi= 0\,.
\ee
The resulting wave equation for the perturbation $\0p$ is
\be
\label{N}
{\0p}\ff - k^2 \e ^{2(2\beta + \psi)}\0p = 0\,.
\ee
The ratio ${\0p}\ff/\0p = k^2\e ^{2(2\beta + \psi)}$ being
regular and positive over $I_\s\/ = ]-\infty,\, 0[$, it is
impossible to keep $\0p$ finite at both ends of $I_{\s}$. Our boundary
conditions (\ref{bc}) cannot be satisfied, so that this case is stable.

Tomimatsu \cite{Tom} investigated the stability of 2-static solutions with
vanishing electric field (\ref{ns}). In the parameter space ($x, y, a$),
he considered, as shown in \cite{thesis}, only the surface portions:
$a = x/2-q,\,y\leq 0,\,x>0$, and $a= x/2+q,\,y>0\,,x<0$, with $y<x^{2}/4$.
He concluded that all the solutions under consideration were unstable,
except the case $y=a=0,\,x\neq 0$ of the Schwarzschild solution which was
found to be stable. The discrepancy with our conclusions can be explained, 
as discussed in \cite{thesis}, by the fact that Tomimatsu treated the
Schr\"{o}dinger eigenvalue problem for an auxiliary function whose direct
physical meaning is not transparent, and used in effect boundary
conditions less stringent than our conditions (\ref{bc}).

\subsection{A set of stable charged solutions}

Let us return to the wave equation (\ref{master}) in the gauge (\ref{j1}).
The ratio $\1p \ff /\1p$ is positive definite over $I_\s$
($\forall\,k^2 >0$) provided the two conditions
\ba
\e ^{2\psi} & \geq & 0\,, \lb{crit1}\\
F \p & \geq & 0\,, \lb{crit2}
\ea
are satisfied simultaneously. Then, if the range $I_\s$ is infinite (which
is the case for $y \le x^2$), and if further $F$ does not have a zero in
this range,
\be\lb{crit3}
F \neq 0 \,, 
\ee 
it is impossible to satisfy our boundary conditions (\ref{bc}), and the
corresponding static solution is stable (a zero of $F$ would lead, as
discussed above, to a pole of the perturbation $\1p(\s)$, which would
enable it to remain finite at both ends of $I_\s$ with $\1p \ff /\1p > 0$
everywhere; as we show in the Appendix the condition (\ref{crit3}) is
always
satisfied when the conditions (\ref{crit1}) and (\ref{crit2}) are
fulfilled). 

Eq. (\ref{sol}) shows that $\e ^{2\psi} = -\lambda_{55}$ changes sign
periodically for $y > x^2/4$, so that condition (\ref{crit1}) cannot be
satisfied, while it is satisfied for $y \le x^2/4$ if $x/2 - a \ge 0$,
implying (on account of (\ref{Q})) 
\be\lb{dom1}
x/2 - a \ge |b|\,. 
\ee

To investigate condition (\ref{crit2}), we compute (for $y < x^2/4$) 
\ba
\psi' & = & \frac{x}{4} + \frac{q}{2}\,\coth (q\s - \eta)\,,\\
\beta' & = & -\frac{x}{2} - \nu\,\coth \nu\s\,,
\ea
where $\eta$ is defined by $q = |b|\sinh \eta$, $x/2 - a = |b|\cosh \eta$
($\eta > 0$). Eq. (\ref{F}) then leads to
\be\lb{F'}
F' = \frac{q^2}{\psi'^2\sinh^2(q\s-\eta)} -
\frac{\nu^2}{\beta'^2\sinh^2(\nu\s)} \,.
\ee
The determination of the range for which this function remains
non--negative is carried out analytically in the Appendix. The conclusion
is
that our stability conditions (\ref{crit1})-(\ref{crit3}) are satified if
either 

a) $x = a - b^2/a$ (implying $y = 0$) with $-2|b| \le a < 0$. This class
of stable solutions, includes black holes ($x > 0$, $a \le 0$) and extreme
black holes ($x = 0$, $a < 0$);

b) $2(a+|b|) \le x < a - b^2/a$ (implying $y > 0$) with $-2|b| \le a <
-|b|$.

\subsection{Charged solutions: two unstable cases, and a stable subcase}

\subsubsection{Case $y>0,\;x=a=0$}

In this massless case ($M = \Sigma = 0$), the matrix $\lambda$ in
(\ref{sol})
reduces to
\be
\lambda = 
\left( \begin{array}{cc}
\cos{b\s} & \sin{b\s} \\
\sin{b\s} & - \cos{b\s}
\end{array} \right) \,,
\ee
leading to a symmetrical Lorentzian wormhole spacetime, with
\be
\e^{2\alpha} = 1\,, \quad \e^{2\beta} = r^2 + \mu^2\,, \quad 
\e^{2\gamma} = \e ^{-2\psi} = (r^2+\mu^2)/(r^2-\mu^2)
\ee
($r = -\mu\cot(\mu\s), \mu = b/2 = Q$).
A deeper investigation of this solution is given in
\cite{geo,stab}. Due to the symmetry of the solution under the change
in the radial coordinate $r \to -r$ and---consequently---of the wave
equation (\ref{master}) (rewritten in terms of the coordinate $r$)
\be
[(r^{2}+\mu^{2})\,{\1p}_{,r}]_{,r}-
\left[k^{2}\,(r^{2}-\mu^{2})+\frac{6\,\mu^{2}}{r^{2}}\right]
\,\1p=0\,,
\ee
the two end-points at spatial infinity ($r \to \pm \infty$) are merged,
leading to only one divergence at $r=+\infty$, which can be cancelled by
choosing $c_2 =0$ in Eq. (\ref{comp2}). At the lower end of the range
[$0,\,+\infty$[ of $r$, the perturbation $\1p$ is unbounded (cf. Eq. (29)
of \cite{stab} and Eq. (\ref{gt5}))
\be\lb{2pole}
\1p = \frac{c_3}{r^{2}}\left[1+
\frac{1}{6}\left(k^{2}+\frac{2}{\mu^{2}}\right)r^{2}+\cdots\right]+
c_4\,r^{3}\left[1-\frac{1}{14}
\left(k^{2}-\frac{34}{3\mu^{2}}\right)r^{2}+\cdots\right]\,.
\ee
Actually the pole in (\ref{2pole}) is gauge dependent and can be removed
by
transforming to the gauge $\delta_2\beta =0$, according to Eq.
(\ref{gt4}), 
which leads to
\ba
\delta_2\psi & = & \frac{r^2}{r^2 -\mu ^2}\,\1p \nonumber \\
& = & -\,\frac{c_3}{\mu^{2}}\left[1+
\frac{1}{6}\left(k^{2}+\frac{8}{\mu^{2}}\right)r^{2}+\cdots\right]
-\,\frac{c_4\,r^{5}}{\mu^{2}}\left[1-\frac{1}{14}
\left(k^{2}-\frac{76}{3\mu^{2}}\right)r^{2}+\cdots\right]\,. \nonumber\\
\ea
This is well bounded for all values of the integration constants. However 
the constraint equation (\ref{R01}) now leads to
\ba\lb{d2a}
\delta_2\alpha\, & = & \,\frac{\delta_2\psi' 
+ (\psi'-\gamma')\delta_2\psi}{2\beta'+\psi'}\, \nonumber \\
& = & \,\frac{c_3 \mu^{2}}{6r^{2}}
\left(k^{2}-\frac{4}{\mu^{2}}\right)+\cdots \,,
\ea
which is singular, unless $c_3(k) = 0$ (leading to an eigenvalue problem),
or 
\be
k=2/\mu\,. 
\ee
We verify
that---for this special eigenvalue ---all the perturbations
$\delta_2\psi$,
etc, are bounded $\forall \, r\,\in\,[0,\,+\infty[\,$. The massless
charged
symmetric wormhole solution is then unstable for it admits a mode of
perturbation growing in time as $\e ^{2t/\mu}$.\\

\subsubsection{Case $y=0,\;a>x>0$}

The functions which appear in Eq. (\ref{master}) are given by
\ba 
\e ^{\beta} & = & \frac{x}{1 - \e ^{x\s}}\,,\lb{ex2}\\ 
\e^{2\psi} & = & \frac{x-a}{x} + \frac{a}{x}\,\e^{x\s}\,,\lb{ex3}\\ 
F & = & \frac{4x-3a}{ax}\,\e^{-x\s} + \frac{3}{x}\,.\lb{ex1}
\ea 
With the
help of Eq. (\ref{y=0}), we see that the third term in Eq. (\ref{master})
vanishes identically if $4x-3a=0$, or behaves at the lower end of 
$I_\s$ ($\s \to -\infty$) as: 
$$ \frac{6F \p}{F^2} \simeq \frac{6ax^2}{3a-4x}\,\e^{x\s}$$ 
if $4x-3a\neq 0$. Since the second term in Eq. (\ref{master}) behaves
as a constant ($\s \to -\infty$): 
$$ \e^{4\beta +2\psi} \simeq - x^3(a-x)\,, $$ 
we can neglect the third term, and write Eq. (\ref{master}) as
\be 
\1p \ff + k^2 x^3 (a-x)\,\1p \simeq 0\,.  
\ee 
We obtain by integration
\be\lb{a>x1} 
\1p \simeq c_{3}\,\cos m\s+c_{4}\,\sin m\s 
\ee 
(with $m=kx\sqrt{x(a-x)}$), which is bounded for all values of the 
integration constants. We must check that the other perturbations
$\delta_1\alpha$ and  $\delta_1\gamma$ are also bounded. 
Using the asymptotic behaviour 
\be\lb{a>x2} 
\frac{\psi \p -\beta \p}{\psi \p +2\beta \p}\;\simeq\;
\left\{\begin{array}{lcl} 
{\ds \frac{x-2a}{3a-4x}} & \rm{if} & 3a-4x\neq
0\\\\ {\ds \frac{1}{2}\,\e^{-x\s}} & \rm{if} & 3a-4x=0\,. 
\end{array}\right. 
\ee
we find from Eqs. (\ref{ta}) and (\ref{tg}) that for $\underline{a \neq
4x/3}$ all the perturbations are finite at $\s \to -\infty$. It follows
that the boundary conditions (\ref{bc}) are satisfied for all values of
$k$ if $c_2 = 0$ in (\ref{comp2}), and the corresponding background
solutions are  unstable.

However if $\underline{a = 4x/3} = 2|b|$ ($\Sigma = -2\sqrt{3}M$, $|Q| =
2M$), $\delta_1\alpha$  and $\delta_1\gamma$ diverge, from (\ref{ta}),
(\ref{tg}) and (\ref{a>x2}), as $\e^{-x\s}\, \delta_1\psi$.  These
divergences may not be removed by transforming to another gauge. For
instance, in the gauge $\delta_2\beta = 0$, $\delta_2\psi \sim
\e^{x\s}\,\delta_1\psi$ from (\ref{gt4}), leading from the first equation
(\ref{d2a}) to a $\delta_2\alpha$ diverging as $\e^{-x\s}$ times a bounded
function. A similar divergence results in a generic linear gauge
($\delta\beta + \lambda\delta\psi = 0$) from the analysis of Eqs.
(\ref{R01}) and (\ref{gt3}). So at least one of the perturbations is never
bounded\footnote{One can show that $\delta\alpha$ is bounded in the gauge
$\delta\alpha - \delta\beta = 0$; however $\delta\gamma$ is not bounded in
this gauge.}, leading to the conclusion that this special subcase is
stable.

\section{Conclusion}

We have discussed the geometry of the 3--parameter family of static
spherically symmetric solutions of 5--dimensional Kaluza--Klein theory. We
have found that, besides the 2--parameter black hole class ($y = 0,\,x >
0,
\,a \le 0$) and the exceptional regular solution ($y = 0,\,a = x > 0$),
this
family contains a 3--parameter class of geodesically complete solutions
($y \ge 3x^2/4,\,x > 0$), which is larger than previously thought. These
regular solutions are necessarily charged.

We have also investigated the stability of these solutions under radial
perturbations. We have shown that all neutral solutions ($b = 0$) are
stable. Among charged solutions, we have found two stability classes ($y =
0,\,a < 0,\,x \ge 3a/4$) and ($0 < y \le x^2/4,\,-2|b| \le a < -|b|$); the
first stablity class includes all the black hole and extreme black
hole solutions. We have also been able to prove analytically the
unstability of two lower--dimensional (in the 3--dimensional parameter
space) classes of solutions. The remaining cases (which include the
geodesically complete solutions) lead to eigenvalue problems. Pending a
numerical investigation of these, we can only conjecture that the
corresponding static solutions are unstable.

This work should be generalized in two directions. Electrostatic
spherically symmetric solutions are only a subcase of the stationary
spherically symmetric solutions which may be obtained from Eq.
(\ref{geotar}), and which are discussed in the extreme case in
\cite{spat,Rasheed}. The stability of these solutions could be
investigated along the lines followed here. Also, the proof of stability
under monopole perturbations is but a first step. The effect of higher
multipole perturbations should also be ascertained before one can conclude
to stability under all small perturbations, but this is a far more
difficult task.

\vspace{0.5cm}
{\bf Acknowledgements:} A large part of this work was carried out while
M.A.-A.  was at the Eastern Mediterranean University, Gazimagusa, North
Cyprus and G.C.  was at the Laboratoire de Gravitation et Cosmologie
Relativistes (LGCR), Paris, France.  G.C. thanks the Universidade Federal
do Esp\'{\i}rito Santo, Vit\'oria, Brasil for kind hospitality during part
of this work.  J.C.F. thanks the LGCR for their hospitality during the
elaboration of part of this work. We thank also CAPES (Brasil) and
COFECUB (France) for financial support through the collaboration project
number 180/96.

\section*{Appendix}
\appendix
\def\theequation{A.\arabic{equation}}
\setcounter{equation}{0}

In this Appendix we determine (for $y < x^2/4$ and $x/2 - a \ge |b|$) the
parameter domains for which the function
\be\lb{F'A}
F'(\s) = \frac{q^2}{\psi'^2\sinh^2(q\s-\eta)} -
\frac{\nu^2}{\beta'^2\sinh^2(\nu\s)} 
\ee
remains non--negative in the range $]-\infty,0]$.  There are 3 cases
according to the value of $y$:\\

1) $y < 0$ ($q > \nu$) . For $\s \to -\infty$, $F' \simeq
-\nu^2/\beta'^2\sinh^2\nu\s$, so that condition (\ref{crit2}) is not
satisfied.\\

2) $y = 0$ ($q = \nu = |x|/2$), i.e. $b^2 = a(a-x)$. Then
\be\lb{y=0}
F' = \frac{3a-4x}{a}\,\e^{-x\s}\,.
\ee
Remembering that $a \le x/2$, we find that condition (\ref{crit2}) is
satisfied for $a < 0$, $x \ge 3a/4$, implying $b^2 \ge a^2/4$. 
Condition (\ref{crit3}) is then also
satisfied because
\be\lb{nopole}
F(\s) \le  F(0) = 2/\psi'(0) = 4/a <  0  \,.
\ee
The corresponding domain of stability is 
\be
x = a - b^2/a\,, \quad -2|b| \le a < 0\,.
\ee

3) $y > 0$ ($q < \nu$). Then, Eq. (\ref{F'A}) can be written as
\be
F' = \frac{4\sinh^2\delta}{\sinh^2(q\s - \eta +\varepsilon\delta)} -
\frac{\sinh^2\kappa}{\sinh^2(\nu\s + \varepsilon\kappa)} \,.
\ee
where $\delta > 0$ and $\kappa > 0$ are defined by 
\ba
& {\ds\frac{x}{2}} = \varepsilon\sqrt{y}\cosh\delta\,, \quad & q =
\sqrt{y}\sinh\delta
\nonumber \\
& {\ds\frac{x}{2}} = \varepsilon{\ds\frac{\sqrt{y}}{2}}\cosh\kappa\,,
\quad & \nu = \frac{\sqrt{y}}{2}\sinh\kappa
\nonumber 
\ea
($\varepsilon = {\rm sign}(x)$). If $x > 0$, $\nu\s + \kappa$ has a zero
for $\s = -\kappa/\nu$, with $F' \to -\infty$, so that condition
(\ref{crit2}) is not satified.

If $x < 0$, $F'(0) = (4b^2 - a^2)/a^2$ is non--negative if  
\be\lb{dom2}
b^2 \ge a^2/4\,.
\ee
$F'(\s)$ can vanish only at the zeroes of
\be
h \equiv f - g \,,
\ee
where the functions 
\be
f \equiv -\sinh(\nu\s - \kappa)\,, \quad g \equiv -\frac{\nu}{q}\sinh(q\s
- \eta - \delta)\,.
\ee
are positive for $\s$ negative. However
$h(\s)$ stays positive in $I_\s$
because the inequalities
\be\lb{app}
(\nu - q)\s < 0 < \kappa - \eta - \delta
\ee
(see below) imply
\be
h' = -\nu[\cosh(\nu\s - \kappa) - \cosh(q\s - \eta - \delta)] < 0\,,
\ee
so that the minimum  value of $h$ is $h(0) =
(\nu/\sqrt{y})(2+a/|b|) \ge 0$. Therefore $F'$ stays positive in $I_\s$,
so that (condition (\ref{crit3}) being again satisfied by virtue of
(\ref{nopole})) the corresponding static solutions, with parameters
\be\lb{dom3}
2(a+|b|) \le x < a - b^2/a\,, \quad -2|b| \le a < -|b|
\ee
(the first two inequalities result from (\ref{dom1}) and $y > 0$, the
last two from (\ref{dom2}) and $x < 0$ together with (\ref{dom1}))
are stable.

It remains to prove the second inequality (\ref{app}). From the
definitions of $\eta$, $\kappa$ and $\delta$ we obtain
\ba
|b|y\sinh(\kappa-\eta-\delta) & = & (\nu - q)ax - 2\nu y \nonumber \\
& = & 2\nu(a^2-b^2) - (\nu+q)ax \\
& \ge & 2(|a|-|b|)(\nu|b| - q|a|)\,,
\ea
using the first inequality (\ref{dom3}). The first factor is non--negative
from the last inequality (\ref{dom3}), while
\be
\nu^2b^2 - q^2a^2 \, = \, (x^2/4)(b^2-a^2) + y(a^2-b^2/4) \,
\ge \, (3b^2/4)y \, > 0
\ee
for $0 < y \le x^2/4$.

\end{document}